\documentclass[prd, aps, reprint, amsfonts, amssymb, amsmath, preprintnumbers, showpacs, superscriptaddress]{revtex4-2}
\usepackage{graphicx, multirow,soul}
\usepackage{enumitem}
\usepackage[citecolor=blue]{hyperref}
\usepackage[all]{hypcap}
\usepackage{url}
\usepackage{color}
\usepackage{subfigure}
\usepackage{slashed}
\usepackage{float}
\usepackage{cancel}
\usepackage{bm}
\usepackage{ulem}
\usepackage{bbm}
\usepackage{bbold}
\usepackage{mathrsfs}
\usepackage{lipsum}
\newcommand{\equaref}[1]{Eq.~(\ref{#1})}

\newcommand{\figref}[1]{Fig.~\ref{#1}}

\newcommand{\appref}[1]{Appendix~\ref{#1}}
\newcommand{\tabref}[1]{Table~\ref{#1}}

\usepackage{tikz}
\tikzset{node distance=2cm, auto}

\newcommand{\bq}{\begin{eqnarray}}
\newcommand{\nq}{\end{eqnarray}}

\newcommand{\be}{\begin{equation}}
\newcommand{\ee}{\end{equation}}
\newcommand{\bea}{\begin{eqnarray}}
\newcommand{\eea}{\end{eqnarray}}
\newcommand{\ba}{\begin{aligned}}
\newcommand{\ea}{\end{aligned}}

\begin{document}

\title{Testing Realistic $SO(10)$ SUSY GUTs with Proton Decay and Gravitational Waves}

\author{Bowen Fu}
\email{B.Fu@soton.ac.uk}
\author{Stephen F. King}
\email{king@soton.ac.uk}
\affiliation{School of Physics and Astronomy, University of Southampton, Southampton, SO17 1BJ, U.K.}
\author{Luca Marsili}
\email{luca.marsili@durham.ac.uk}
\affiliation{Institute for Particle Physics Phenomenology, Department of Physics, Durham University, Durham DH1 3LE, U.K.}
\author{Silvia Pascoli}
\email{silvia.pascoli@unibo.it}
\affiliation{DIFA, University of Bologna, via Irnerio 46, 40126 Bologna, Italy}
\affiliation{INFN, Sezione di Bologna, viale Berti Pichat 6/2, 40127 Bologna, Italy}
\author{Jessica Turner}
\email{jessica.turner@durham.ac.uk}
\affiliation{Institute for Particle Physics Phenomenology, Department of Physics, Durham University, Durham DH1 3LE, U.K.}
\author{Ye-Ling Zhou}
\email{zhouyeling@ucas.ac.cn}
\affiliation{School of Fundamental Physics and Mathematical Sciences, Hangzhou Institute for Advanced
Study, UCAS, Hangzhou 310024, China}
\affiliation{International Centre for Theoretical Physics Asia-Pacific, Beijing/Hangzhou, China}


\begin{abstract}
We present a comprehensive analysis of a supersymmetric $SO(10)$ Grand Unified Theory, which is broken to the Standard Model via the breaking of two intermediate symmetries. 
The spontaneous breaking of the first intermediate symmetry, $B-L$, 
 leads to the generation of cosmic strings and right-handed neutrino masses and further to an observable cosmological background of gravitational waves and generation of light neutrino masses via type-I seesaw mechanism. 
 Supersymmetry breaking manifests as sparticle masses below the $B-L$ breaking but far above the electroweak scale due to proton decay limits. 
 This naturally pushes the $B-L$ breaking scale close to the GUT scale, leading to the formation of metastable cosmic strings, which can provide a gravitational wave spectrum consistent with the recent Pulsar Timing Arrays observation. 
 We perform a detailed analysis of this model using two-loop renormalisation group equations, including threshold corrections, to determine the symmetry-breaking scale consistent with the recent Pulsar Timing Arrays signals such as NANOGrav 15-year data and testable by the next-generation limits on proton decay from Hyper-K and JUNO. 
 Simultaneously, we find the regions of the model parameter space that can predict the measured quark and lepton masses and mixing, baryon asymmetry of our Universe, a viable dark matter candidate and can be tested by a combination of neutrinoless double beta decay searches and limits on the sum of neutrinos masses. 
\end{abstract}
\preprint{IPPP/23/41} 
\maketitle

\section{Introduction}\label{sec:intro}

$SO(10)$ Grand Unified Theories (GUTs) are widely studied, ultraviolet complete frameworks that unify three of the fundamental forces and have unique features \cite{Fritzsch:1974nn,Georgi:1975qb,Georgi:1974my}:
\begin{itemize}[noitemsep,topsep=0pt]
\item 
Highly correlated fermion masses and mixing as quarks and leptons are arranged in a single representation in the GUT gauge space.
\item 
Inclusion of a $U(1)_{B-L}$ gauge symmetry, whose spontaneous breaking gives raise to cosmic strings \cite{Kibble:1982ae} and right-handed neutrino masses, that can generate light Majorana masses for neutrinos and a baryon asymmetry. 
\item 
Generically, baryon and lepton number violating operators that induce proton decay and allow to set stringent limits on the GUT breaking scale.
\end{itemize}
These features of $SO(10)$ GUTs allow to constrain the models with present and future data from a variety of complementary approaches.
Next-generation large-scale neutrino experiments, including JUNO~\cite{JUNO:2015zny}, DUNE \cite{DUNE:2015lol}, and Hyper-Kamiokande \cite{Hyper-Kamiokande:2018ofw}, are expected to measure the majority of neutrino oscillation parameters at percent level precision. An additional goal of these experiments will be to constrain, or possibly even measure, the proton lifetime at an unprecedented level. Both measurements will probe the theory parameter space of $SO(10)$ GUTs. Moreover, as discussed above, a generic feature of many GUTs is the breaking of a $U(1)$ gauge symmetry which can generate a network of cosmic strings \cite{Kibble:1976sj}. If the string network is not completely diluted by inflation, it may be a source of stochastic gravitational wave background (SGWB) with a broad spectrum of frequency from nanohertz to kilohertz. Due to this broad spectrum, a variety of currently running and upcoming gravitational wave (GW) experiments, including those from Pulsar Timing Arrays (PTAs), space- and ground-based laser interferometers, as well as atomic interferometers, will be able to constrain such GUTs \cite{Buchmuller:2013lra,Dror:2019syi,Buchmuller:2019gfy,Chigusa:2020rks,Saad:2022mzu}. Very recently, the NANOGrav collaboration with 15 years of data activity (NANOGrav15) \cite{NANOGrav:2023gor,NANOGrav:2023icp,NANOGrav:2023hfp, NANOGrav:2023ctt,NANOGrav:2023hvm} reported the evidence for quadrupolar correlations \cite{Hellings:1983fr}, indicating a GW origin of the signal. Similar evidences have been independently reported by EPTA \cite{EPTA:2023fyk,EPTA:2023sfo,EPTA:2023akd,EPTA:2023gyr,EPTA:2023xxk,EPTA:2023xiy}, PPTA ~\cite{Zic:2023gta,Reardon:2023zen,Reardon:2023gzh}, and CPTA~\cite{Xu:2023wog}. 
Based on the Bayesian analysis of NANOGrav15, it was shown that the Nambu-Goto (NG) strings do not provide a good fit to the signal \cite{NANOGrav:2023hvm} and sets a stringent bound on the symmetry breaking scale for stable strings. The possibility that metastable cosmic strings \cite{Buchmuller:2023aus,Antusch:2023zjk,Madge:2023cak} or superstrings \cite{Ellis:2023tsl} may be the source of the observed SGWB remains open. 

We have shown that this rapid progress in neutrino and gravitational wave measurements provides complementary probes of $SO(10)$ GUTs \cite{King:2020hyd,King:2021gmj}, offering a unique opportunity to indirectly test very high energy scales. In Ref.~\cite{Fu:2022lrn}, we carried out a detailed study of $SO(10)$, showing that all Standard Model (SM) fermion masses and mixing parameters and the baryon asymmetry can be matched to their observed values in this model. In that work, we constructed a model with a $U(1)_{B-L}$ symmetry breaking scale around $10^{13}$~GeV, that is not excluded by PTAs but cannot explain the newer indications of SGWB that may originate from metastable strings.

In this Letter, we continue our roadmap on the testability of $SO(10)$ GUTs by extending the analysis to a supersymmetric (SUSY) version. As we will show, SUSY GUT can have marked differences with respect to the non-SUSY models considered already. It offers a natural framework for metastable strings consistent with the NANOGrav signal, as it favours a $U(1)_{B-L}$ breaking very close to the GUT scale.
In addition to being currently one of the favoured explanation to the SGWB signal recently observed by the PTA observatories, thanks to its rather broad frequency spectrum the resulting SWGB will be within the sensitivity of future GW experiments at higher frequencies.
Moreover, one of the key predictions of SUSY GUTs is kaonic proton decay ($p \to K \bar{\nu}$), in addition to the other non-SUSY decay channels. In the coming years, JUNO will provide the best sensitivity to this channel and place a key constraint on SUSY GUTs. 

We perform a comprehensive analysis of this model by determining each scale of symmetry breaking by solving the renormalisation group equations (RGEs) and fitting our model to SM fermion masses and mixing data. 
We concretely demonstrate that the predictions of our model can be tested by next-generation cosmic microwave background (CMB) observations, neutrinoless double beta decay, oscillation measurements, GW experiments, and searches for proton decay. Moreover, our model can predict the observed baryon asymmetry and accommodate a viable dark matter candidate. 

\section{Model Framework \label{sec:model}}

The model we present and confront with flavour, proton decay, and gravitational wave data is
a SUSY $SO(10)$ GUT, that is spontaneously broken to the SM as follows: 
\bea\label{eq:breaking}
& SO(10) \times \text{SUSY} \nonumber\\
&\hspace{5mm}{\bf 45}~\Big\downarrow~\text{broken at} ~ M_{\rm GUT} \nonumber\\
&G_{\rm LRSM} \equiv SU(3)_c \!\times\! SU(2)_L \!\times\! SU(2)_R \!\times\! U(1)_{B-L} \!\times\! \text{SUSY} \nonumber\\
&\hspace{3mm}{\overline{\bf 126}}~\Big\downarrow~\text{broken at} ~ M_{B-L} \nonumber\\ 
&G_{\rm MSSM} \equiv SU(3)_c \times SU(2)_L \times U(1)_Y \times \text{SUSY} \nonumber\\
&\hspace{10mm}{}~\Big\downarrow~\text{broken at} ~ M_{\rm SUSY} \nonumber\\ 
&G_{\rm SM} \equiv SU(3)_c \times SU(2)_L \times U(1)_Y \,.
\eea
At scale $M_{\rm GUT}$, the GUT symmetry is broken, and dimension-six operators which mediate proton decay are induced. This GUT symmetry breaking also leads to the production of monopoles which we assume are removed through a period of rapid inflation. 
The next breaking step occurs at $M_{B-L}$, where the $U(1)_{B-L}$ gauge symmetry is spontaneously broken. This leads to the production of a network of cosmic strings, that can decay gravitationally, as well as to the generation of right-handed neutrino (RHN) masses. These RHNs can decay to produce a matter-antimatter asymmetry via thermal leptogenesis \cite{Fukugita:1986hr}.
In the final step, SUSY is broken at a scale $M_{\rm SUSY}$, defined to be equal to the common masses of the squarks and sleptons, and dimension-five operators which mediate proton decay are induced via wino and higgsino exchange, whose masses, $M_{\widetilde{W}}$, are allowed to be below $M_{\rm SUSY}$. This is typical of widely studied split SUSY scenarios \cite{Giudice:2004tc,Arkani-Hamed:2004ymt}.
The boldface numbers to the left of the arrows of \equaref{eq:breaking} denote the {representations of Higgs superfields} of $SO(10)$ required for the symmetry breaking. 

\begin{figure}[t!]
 \centering
 \includegraphics[width = 0.48\textwidth]{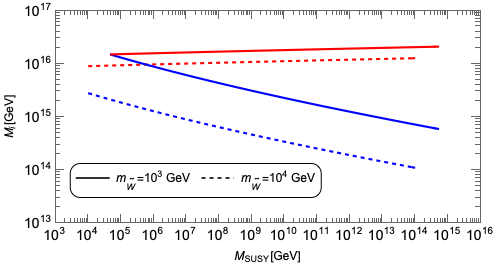}
 \caption{The GUT scale, $M_{\rm GUT}$ (red), and $U(1)_{B-L}$ breaking scale, $M_{B-L}$ (blue), as a function of the SUSY breaking scale $M_{\rm SUSY}$, defined as the common squark and slepton mass scale. The solid and dashed lines show the effect on the RGEs solutions for various wino masses, assumed to be below $M_{\rm SUSY}$. For an observable GW signal, the inflationary scale must lie between the red and blue lines.} 
 \label{fig:running}
\end{figure}

In Ref.~\cite{King:2021gmj}, we analysed all symmetry-breaking patterns of a non-SUSY $SO(10)$ GUT to the SM via the Pati-Salam path by solving the two-loop RGEs and assuming gauge coupling unification at $M_{\rm GUT}$. We found that the majority of breaking patterns constrain $M_{B-L}\lesssim 10^{13}$ GeV with no symmetry breaking patterns attaining $M_{B-L} \gtrsim 10^{14}$ GeV. This leads to the formation of cosmic strings with a relatively small string tension which is more difficult to be tested by GW interferometers. One notable motivation to extend from non-SUSY GUTs to SUSY GUTs is that, in the SUSY version, $M_{B-L}$ can naturally reach $10^{14}-10^{15}$ GeV, which leads to the formation of very heavy metastable strings that can accommodate the GW signal detected by the PTAs. To determine the $B-L$ breaking scale in the SUSY version, we follow the same procedure as our previous work Ref.~\cite{King:2021gmj}, by solving the two-loop RGEs, including threshold effects from gauginos and higgsino masses, which may be a few orders of magnitude lower than $M_{\rm SUSY}$. The $\beta$ coefficients at each intermediate scale are listed in \appref{app:gauge_unification}. From our RGE analysis, we find that the lower the scale of SUSY breaking, the higher the $U(1)_{B-L}$ symmetry breaking scale as shown in \figref{fig:running}. 
As the wino mass increases, the $B-L$ scale is suppressed while the GUT scale remains roughly the same, enlarging the hierarchy between the two scales, that leads to a stable network of strings.
We use the RGE solutions as input for determining the proton decay rate and gravitational wave signal. In the following, we discuss the testability of this model.

\subsection{Fermion masses and mixing angles}
At the GUT scale, the Yukawa superpotential is given by:
\bea\label{eq:yuk}
W_{\rm Y} &= &Y_{\bf 10}^* \, {\bf 16} \cdot {\bf 16} \cdot {\bf 10} + Y_{\overline{\bf 126}}^* \, {\bf 16} \cdot {\bf 16} \cdot \overline{\bf 126} \nonumber\\
&&+Y_{\bf 120}^* \, {\bf 16} \cdot {\bf 16} \cdot {\bf 120} + {\rm h.c.}\,,
\eea
where the asterisk denotes complex conjugation, ${\bf 16}$ is the $SO(10)$ matter multiplet and ${\bf 10}$, $ \overline{\bf 126}$ and ${\bf 120}$ are the Higgs superfields. See \appref{sec:mattercontent} for more details. In flavour space, the Yukawa matrices are $3\times 3$ with $Y_{\bf 10}$ and $Y_{\overline{\bf 126}}$ complex, symmetric and $Y_{\bf 120}$ {real,} antisymmetric. 
We treat the quark masses and CKM mixing parameters \cite{ParticleDataGroup:2022pth,FlavourLatticeAveragingGroupFLAG:2021npn}  as inputs and hence this model has seven free parameters which we vary to predict eight observables in the lepton sector. 
We perform the procedure using {\sc MultiNest} \cite{Feroz:2008xx} to minimise the $\chi^2$ statistical measure as detailed in \appref{sec:matter}..
\subsection{Leptogenesis and $0 \nu \beta \beta$ decay}
  \begin{figure}[t!]
   \centering
   \includegraphics[width = 0.50\textwidth]{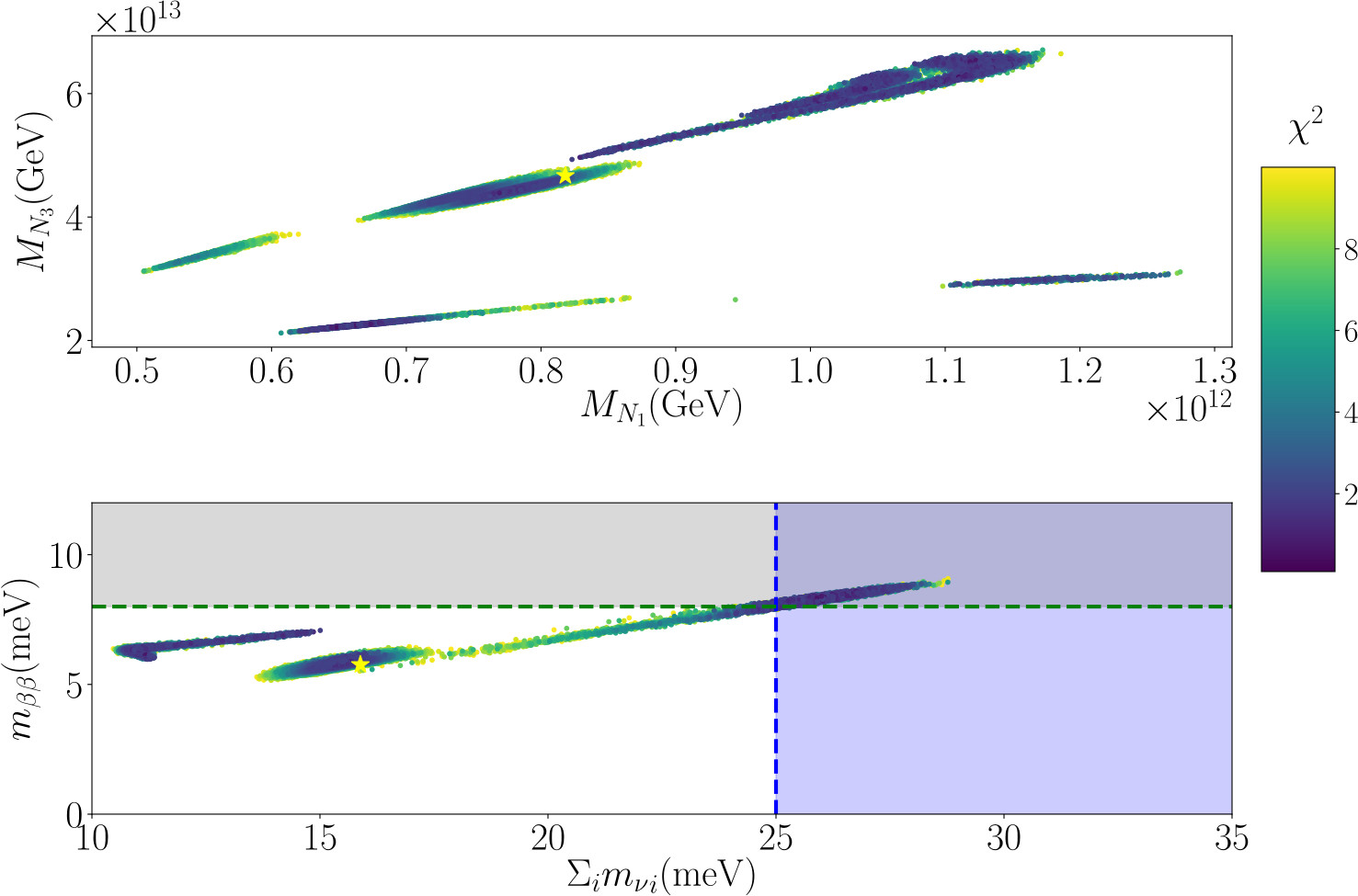}
   \caption{All coloured points show regions of the parameter space that fit the neutrino sector with $\chi^2 < 10$. The yellow star shows a benchmark point in the scan with $\eta_B \sim 6.2\times 10^{-10}$. All the points in the scan predict $-10\lesssim \log_{10}(\eta_B)\lesssim-8$. 
   In the upper plot, we see that the heaviest right-handed neutrino mass is predicted to be $M_{N_{3}}\sim 10^{13} {\rm GeV}$, with the lightest right-handed neutrino mass $M_{N_{1}}$ an order of magnitude less, and the mass hierarchy is mild, with the lightest right-handed neutrino mass $M_{N_{1}}$ an order of magnitude less. In the lower plot, we show the effective neutrino mass ($m_{\beta\beta}$) predicted from our model as a function of the sum of neutrino masses. 
The grey regions shows the reach of the next generation $\nu 0 \beta \beta$ experiments and the blue region shows the sensitivity for EUCLID \cite{Audren:2012vy,Chudaykin:2019ock}.
   \label{fig:running2}}
  \end{figure}
For the points in the model parameter space scan which fits the flavour data at high statistical significance, there is a prediction for the Yukawa matrix, $Y_\nu$, which couples the RHNs to the leptonic and Higgs doublets and the mass spectrum for the RHNs where the latter is shown in the upper panel of \figref{fig:running2}. The heaviest right-handed neutrino mass is constrained at around $M_{N_{3}}\sim10^{13}\,\rm {GeV}$ and we observed that there is a mild mass hierarchy predicted. We can calculate the baryon asymmetry generated from the decays of these RHNs using {\tt ULYSSES} \cite{Granelli:2020pim,Granelli:2023vcm} to solve the density matrix equations. All points of the scan which fit the flavour data allow for viable leptogenesis with a baryon asymmetry of the same order or larger than the observed value \cite{Planck:2018vyg}.
We found that the model parameter space highly favours normally ordered neutrino masses, with the lightest neutrino mass in the range $5\lesssim m_{\nu_1}\,({\rm meV})\lesssim15$. In the lower panel of \figref{fig:running2}, we show the predictions of our scan for the effective Majorana mass ($m_{\beta\beta}$) as a function of the sum of neutrino masses ($\sum_{i} m_{\nu_{i}}$). Both observables are testable by the next generation of $\nu 0 \beta \beta$ \cite{LEGEND:2021bnm,nEXO:2021ujk, DARWIN:2020jme, SNO:2015wyx, Armengaud:2019loe} and CMB experiments \cite{Audren:2012vy,Chudaykin:2019ock,Boyle:2018rva} with their sensitivities shown in grey and blue, respectively.
We note that all the points shown fit the flavour data well ($\chi^2\lesssim 10$), and our benchmark point (indicated by the associated yellow star in \figref{fig:running2} see \appref{sec:BM} for the Yukawa matrices) is consistent with the flavour data ($\chi^2\lesssim3$) and provides a prediction of the baryon-to-photon ratio $\eta_B\sim 6.2\times 10^{-10}$. \\

\subsection{Proton Decay}
The proton decay bound on the GUT scale can be translated to the restriction on the SUSY breaking scale and wino mass 
from the solutions of the RGEs (see \figref{fig:running}) and the assumption of gauge coupling unification. 
Increasing $m_{\widetilde{W}}$ suppresses $M_{\rm GUT}$, and therefore increases the proton decay rate to a level which can be tested in Hyper-K, as later summarised in Fig.~\ref{fig:ConandSen}. Such low mass sparticles may be within reach at the FCC \cite{FCC:2018byv}, offering another avenue to test the model.
As the GUT we study is supersymmetric, additional contributions to proton decay from the colour-triplet Higgs superfields can mediate the baryon-antilepton transition. 
The consequence is that some decay channels, such as $p \to K^+ \bar{\nu}$, are enhanced. 
Although this channel is less well experimentally constrained, $\tau_{K\bar{\nu}} \gtrsim 5.9\times 10^{33}$ years in Super-K \cite{Super-Kamiokande:2014otb}, the SUSY GUT provides $\tau_{K\bar{\nu}} \propto M_{\rm GUT}^2 M_{\rm SUSY}^2$, and thus this channel can lead to stronger constraints on the SUSY GUT. Since we consider split SUSY spectrum, there is an additional enhancement to the partial lifetime by a factor $M_{\rm SUSY}^2/m_{\widetilde{W}}^2$ in the case $m_{\widetilde{W}} \ll M_{\rm SUSY}$. Moreover, as this channel originates from the Yukawa superpotential in \equaref{eq:yuk}, the partial lifetime is determined by the Yukawa coupling matrices, which are almost entirely fixed, up to overall order-one factors, by our fit to the fermion flavour data. Further details of the proton lifetime calculations are provided in \appref{app:proton_decay}. 
We will discuss how the pionic and kaonic decay channels can constrain the GUT model parameter space and the non-trivial interplay with the GW predictions in the discussion section.

\subsection{Dark Matter}
If R-parity is conserved after SUSY breaking, the lightest SUSY particle (LSP) would be stable and thus can be a dark matter candidate if it is a neutralino \cite{Low:2014cba,Fukuda:2018ufg}. Due to the observed relic abundance, an upper limit on the mass of LSP can be obtained to avoid over-abundance. 
Generally, the lightest neutralino can be the mixture of wino, bino and higgsino. The maximal dark matter mass can be as high as 10 TeV with resonant heavy Higgs annihilation or enhancement in annihilation rate through next-to-LSP \cite{Profumo:2005xd}.
On the other hand, a pure bino LSP or bino-dominated LSP commonly leads to an over-abundance unless (co)annihilation processes further reduce the relic density. The upper limit of pure wino dark matter mass is around 3 TeV \cite{Cohen:2013ama}, and for pure higgsino the limit is around 1 TeV \cite{Giudice:2004tc} which can account for the correct dark matter relic density.

\subsection{Gravitational Waves} 

When a $U(1)$ symmetry is broken at a certain scale $M_{B-L}$, a cosmic strings network is generated. 
The long strings in the network can intersect to form loops that oscillate and emit energy via gravitational radiation.
Such radiation is not coherent and can be seen as a stochastic background of gravitational waves.
Importantly, this background can, in principle, be observed by currently running and future GW experiments. 

We begin with the Nambu-Goto string approximation, where the string is infinitely thin with no couplings to particles \cite{Vachaspati:1984gt}, and the amplitude of the relic GW density parameter is:
\be
\Omega_{\rm GW} (f) = \frac{1}{\rho_c} \frac{d\rho_{\rm GW}}{d \log f}\,,
\ee
where $\rho_c$ is the critical energy density of the Universe and $\rho_{\rm GW}$ depends on a single parameter, $G\mu$ where $G= M_{\rm pl}^{-2}$ is Newton's constant and 
$\mu$ is the string tension. For strings generated from the gauge symmetry $G_{\rm int} = SU(3)_c \times SU(2)_L \times SU(2)_R \times U(1)_{B-L}$, $G\mu$ is approximately given by \cite{Vilenkin:1984ib}:
\begin{eqnarray}
G\mu \simeq 
\displaystyle\frac{1}{2(\alpha_{2R}(M_{B-L})+\alpha_{1X}(M_{B-L}))} \frac{M_{B-L}^2}{M_{\rm pl}^2}\,.
\end{eqnarray}
In the case that $M_{B-L}$ is not far away from $M_{\rm GUT}$, deviation of $\alpha_{2R}(M_{B-L})$ and $\alpha_{1X}(M_{B-L})$ from $\alpha_{\rm GUT}(M_{\rm GUT})$ due to RG running is small. So we can approximate
\begin{eqnarray}
G\mu \simeq 
\displaystyle\frac{1}{4\alpha_{\rm GUT}(M_{\rm GUT})} \frac{M_{B-L}^2}{M_{\rm pl}^2}\,.\label{eq:gmu}
\end{eqnarray}
Moreover, hence we can relate the string tension parameter to the intermediate scale $M_{B-L}$.

As the model we consider has $U(1)$ breaking energy scale close to the GUT symmetry breaking (which generates monopoles), the cosmic strings network can decay via monopole-antimonopole nucleation as studied in Refs.~\cite{Buchmuller:2019gfy,Buchmuller:2020lbh,Masoud:2021prr}.
After the stage of string formation, the cosmic strings network and the consequent loops start to decay producing monopoles-antimonopoles pairs. The exponential suppression of the loop number density is characterised by the decay width per unit length $\Gamma_d$, 
\begin{eqnarray}
\Gamma_d = \frac{\mu}{2 \pi} e^{-\pi \kappa}\,, \quad \kappa = \frac{m^2}{\mu}\,,
\end{eqnarray}
for strings and the consequent loops, where $m = \frac{M_V}{\alpha} f_{\rm m}$ the monopole mass, and
$f_{\rm m}$ is an undetermined factor depending on the model's detail usually assumed of order one \footnote{While the factor $f_{\rm m}$ in the monopole mass is assumed to be of order one \cite{Preskill:1984gd}, it is worth noting that $f_{\rm m} \to 0$ could happen in the case of nonmaximal symmetry breaking \cite{Weinberg:2006rq}. In the latter case, $M_{\rm GUT} \gg m_{\rm m} \sim \sqrt{\mu}$ is still allowed. Thus, we should remember that $M_{B-L} \ll M_{\rm GUT}$ could still be achievable when fitting the NANOGrav data for some special SUSY GUT models}.
In practice, it translates into a cutoff in the low-frequency spectrum of the gravitational waves background which avoids the region tested by Pulsar Time Arrays and provides the appropriate tilt in the spectrum observed by the PTA experiments. These signals can be tested by high-energy gravitational waves experiments such as the Einstein Telescope and, at high energies, by LIGO-Virgo-KAGRA. 

Although the stable cosmic strings are disfavoured, the source of the signal detected by the PTAs \cite{NANOGrav:2023gor,NANOGrav:2023hvm,EPTA:2023fyk,Reardon:2023gzh,Xu:2023wog} may be a network of metastable cosmic strings, which occurs when
 the string network decays to monopole-antimonopole pairs \cite{Vilenkin:1982hm,Preskill:1992ck,Leblond:2009fq,Monin:2008mp,Buchmuller:2021mbb}. The hierarchy between the GUT scale and the string formation scale can be parametrised by 
 \be 
\sqrt{\kappa} \simeq \alpha_{\rm GUT}^{-1/2}\, \frac{M_{\rm GUT}}{M_{B-L}}\,,
 \label{eq:kappa}
\ee
where an order-one coefficient is ignored. The smaller $\kappa$, the closer the GUT and string scales and the more efficient the annihilation of the string network. As we have shown, this SUSY $SO(10)$ GUT prefers a small hierarchy between $M_{\rm GUT}$ and $M_{B-L}$, which naturally leads to a prediction of metastable cosmic strings.
Metastable strings have been suggested as a possible explanation of the PTA observations which favour $\sqrt{\kappa}\approx 8$. We can use the PTA observations as one of the strongest constraints for a stable network of cosmic strings requiring that $G\mu < 2\times 10^{-10}$ \cite{NANOGrav:2023hfp} when $\sqrt{\kappa} \gg 9$, which corresponds to $M_{B-L} \lesssim 6\times 10^{13}\,\text{GeV}$ due to \equaref{eq:gmu}.

By assuming gauge unification, we can directly connect the intermediate scale of the GUT symmetry breaking with the SUSY breaking scale, which we assume to be of the same order of magnitude as the sfermions masses. 
Therefore, we can constrain the SUSY breaking scale using gravitational waves. 
It is worth noticing that the link between the SUSY breaking scale and $M_{B-L}$ depends on the mass of the gauginos, which, as we can see in \figref{fig:running}, heavily affects the running of the gauge coupling.

\subsection{Results and Discussion}

\begin{figure}[t!]
 \centering
 \includegraphics[width = 0.48\textwidth]{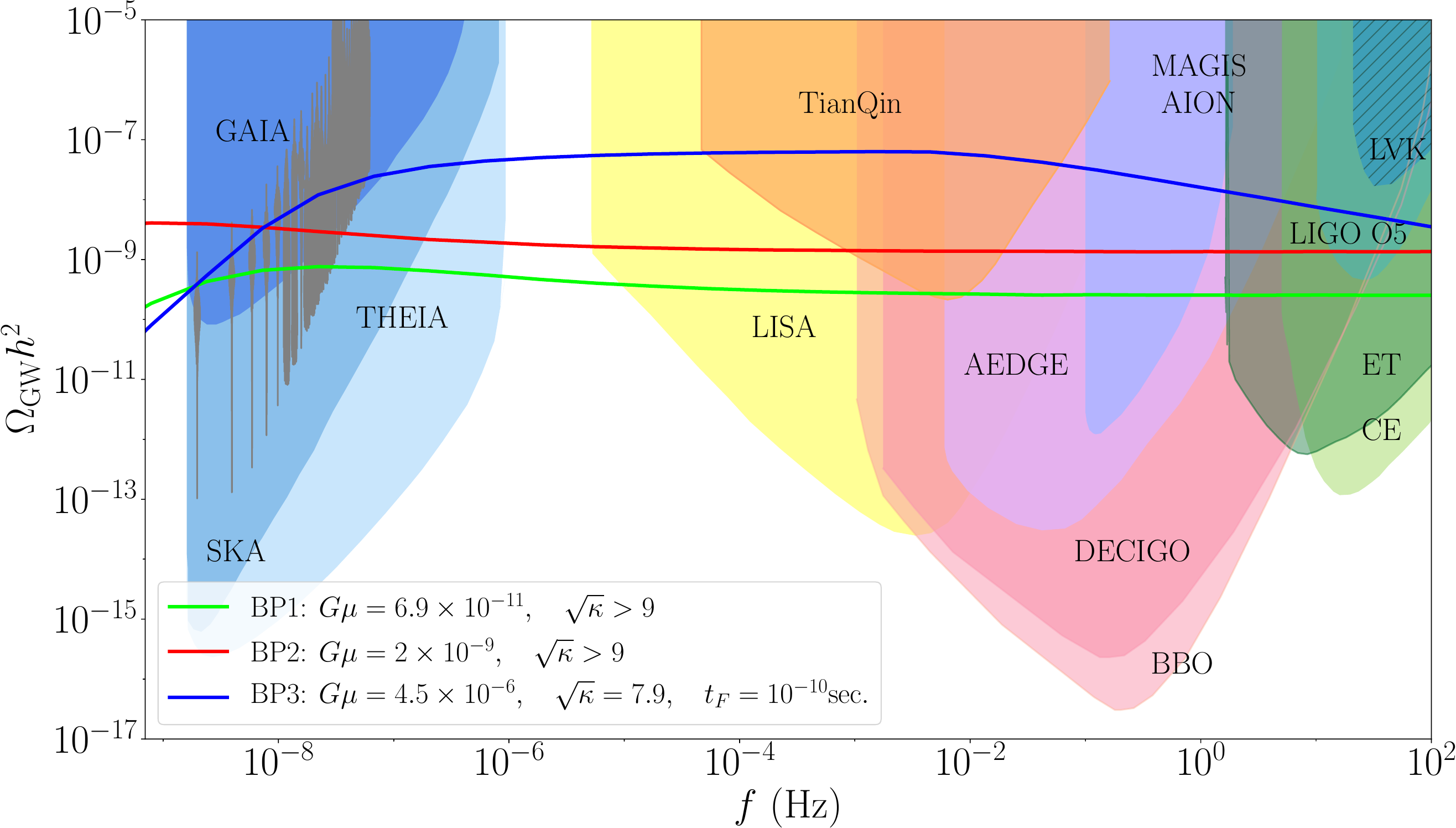}
 \caption{GW spectra of the benchmark points. The model naturally accommodates a signal generated by a partially inflated metastable network (BP3, blue), which supports NANOGrav15. 
 BP1 (BP2) shown in green (red) predicts stable strings which is consistent (inconsistent) with the PTA observations.}
 \label{fig:GUTNANO}
\end{figure}

Here, we consider the various GUT observables and their interplay and focus on three benchmark points, representative of the three key behaviours of the model:
\begin{itemize}
    \item[BP1] This has a high SUSY-breaking scale, ($M_{\rm{SUSY}}\sim 10^{9}$ GeV), as well as wino masses, leading to a $B-L$ breaking scale which is lower than BP2 and BP3. The model exhibits characteristics very similar to the non-SUSY SO(10) GUT.     
    \item[BP2] The SUSY and wino mass scales are still quite high but lower than BP1, and a prediction for proton decay via the SUSY channel $p\rightarrow K \nu$ can be achieved close to current bounds. Therefore, this case could be differentiated from non-SUSY $SO(10)$.
    \item[BP3] This is the most characteristic case for this model. Thanks to the low wino mass and relatively low SUSY mass scale, the $B-L$ breaking scale is very close to $M_\mathrm{GUT}$ leading to the possible generation of metastable strings and a viable explanation of NANOGrav15. A sizable prediction for the SUSY proton decay channel also emerges. 
\end{itemize}
In \figref{fig:GUTNANO}, we study in detail the GW predictions in the three cases, confronting them with NANOGrav15 and the LIGO-Virgo-KAGRA (LVK) bounds \cite{KAGRA:2021kbb} on the spectrum at the high-frequency band. 
For BP1, the green curve shows the GW spectrum from stable strings,
consistent with all constraints, including the upper bound set by PTA. We note that this benchmark does not provide an explanation of the PTA observation. 
For BP2, the red curve shows a GW spectrum from stable strings with a larger $G\mu$ value, 
that is inconsistent with the PTA observations, showing the importance of GW observations in constraining the model (this assumes a sufficiently high inflationary scale).
Finally, for BP3, the blue curve shows the GW spectrum from metastable strings diluted by the inflation in the high-frequency band, which fits NANOGrav15 very well and is testable with future GW observations. 
It is important to note that since the generation of a metastable string network requires $M_{B-L} \sim M_{\rm GUT}$, without any other assumptions, this model tends to favour a signal that would be excluded by the current observations from LVK \cite{KAGRA:2021kbb}. Therefore, to provide an explanation of the PTA observations, we require the cosmic string network to be partially diluted by inflation \cite{Guedes:2018afo,Antusch:2023zjk}, thereby suppressing the signal in the higher frequency regime. 
Since  $M_{B-L}$, $M_{\rm GUT}$, and inflation are at approximately the same scale in such a scenario, it is plausible to have the cosmic string network generated towards the end of inflation and, therefore, are slightly inflated away with a typical $1/f$ suppression in the region in the high-frequency regime, as shown in \figref{fig:GUTNANO}, for a typical time $t_F \sim 10^{-10}\,\rm{s}$.

\begin{figure}[h!]
 \centering
 \includegraphics[width = 0.48\textwidth]{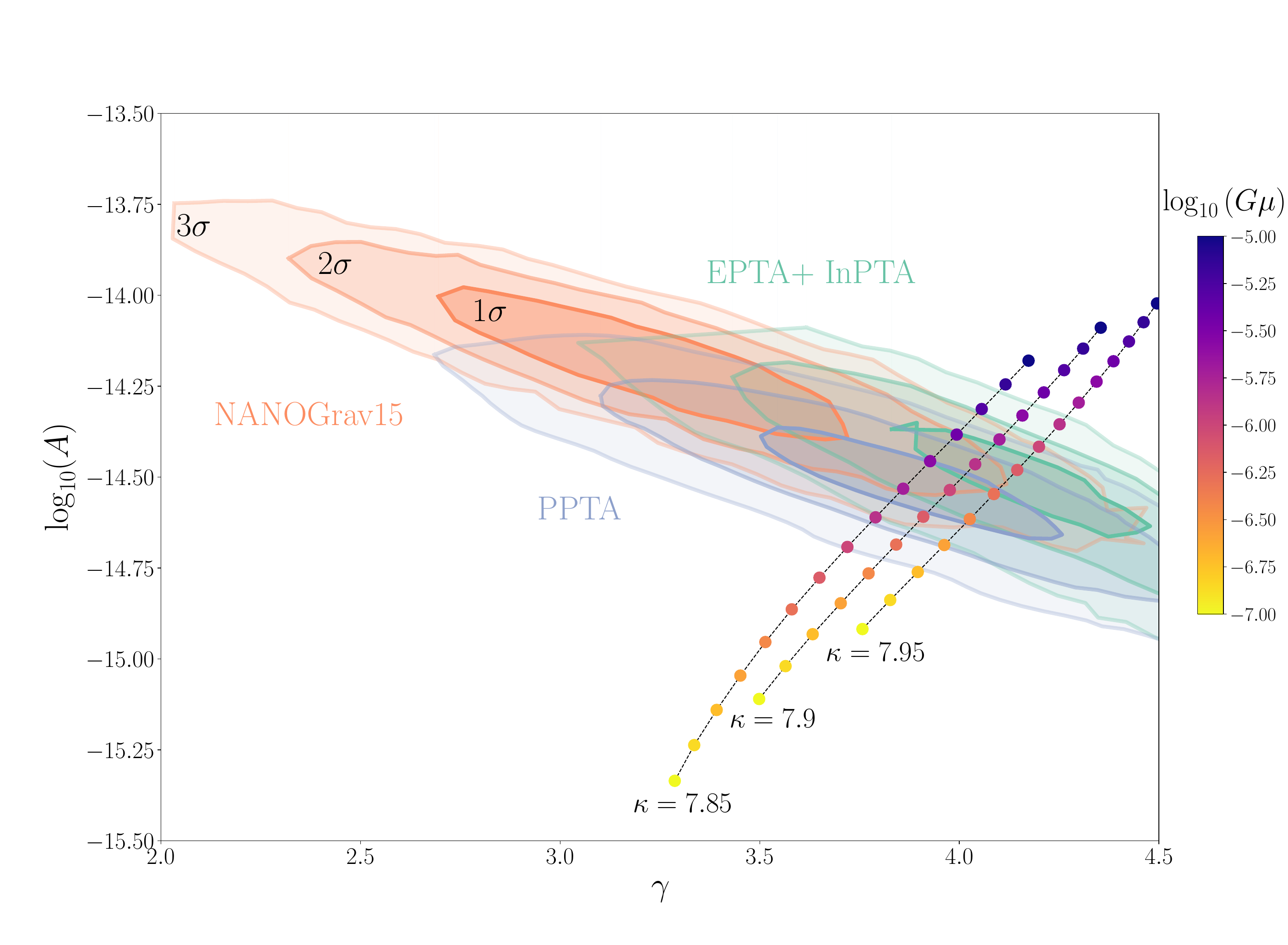}
 \caption{Comparison between SGWB produced by a metastable network of cosmic strings with $G \mu$ from $10^{-7}$ to $10^{-5}$ and three different values of $\kappa$, and the $1 \sigma$, $2\sigma$, $3\sigma$ regions of the signal detected by NANOGRAV \cite{NANOGrav:2023gor}, EPTA \cite{ EPTA:2023fyk}, and PPTA \cite{Reardon:2023gzh}.  }
 \label{fig:Agammacomparison}
\end{figure}

Assuming a power-law spectrum of the characteristic GW strain, the GW energy density spectrum can be parameterised by two parameters: the amplitude parameter, $A$, and power parameter, $\gamma$.
Indeed, the characteristic strain of the signal detected by Pulsar Time Arrays can be parameterised as
\be
h_c(f) = A \left( \frac{f}{f_{\rm yr}} \right)^{\gamma},
\ee
and the GW energy density spectrum , in the nHz region, is given by \cite{Ellis:2020ena}
\be
\Omega(f) = \Omega_{yr} \left(\frac{f}{f_{\rm yr}} \right)^{5-\gamma},
\ee
where $\Omega_{yr} = \frac{2 \pi^2}{3 H_0^2} A^2 f_{\rm yr}^2$. 
We perform the fit in the interval $2-59$ nHz following the procedure of Refs. \cite{Ellis:2020ena, Fu:2022lrn}, and the results are shown in \figref{fig:Agammacomparison}, where the value of $A$ and $\gamma$ favoured by observation is compared with the prediction of the metastable cosmic string network. The reference frequency of the PTA results is chosen to be $1\mathrm{yr}^{-1}$. We found that the signal produced by a metastable network of cosmic strings is compatible with the EPTA and PPTA in the $1\sigma$ range and with NANOGrav in the $2\sigma$ region.
From \figref{fig:Agammacomparison} we note that the predicted signal is very sensitive to the value of $\kappa$. Further, the values of $G \mu$ consistent with the observations from the Pulsar Time Array observatories get higher as $\kappa$ becomes smaller. For the values of $\kappa$ we have shown, $\kappa = 7.85, 7.9, 7.95$, the values of $G \mu$ for which the predicted signal is within the $2 \sigma$ region, are between $10^{-5}$ and $10^{-7}$, depending on $\kappa$, and they are consistent to the scenario in which $M_{B-L} \simeq M_{\rm GUT}$.

\begin{figure}[h!]
 \centering
 \includegraphics[width = 0.48\textwidth]{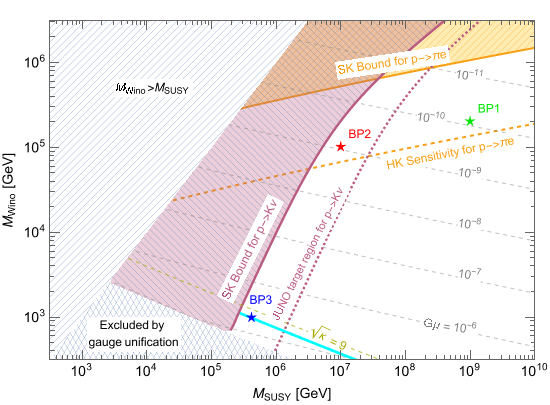}
 \caption{Constraints and sensitivities on proton decay in the $M_{\rm SUSY}-M_{\rm Wino}$ plane,
 (where $M_{\rm SUSY}$ is defined as the squark and slepton mass scale). 
 The orange lines show the constraint of Super-K (solid) and sensitivity of Hyper-K (dashed) on the $p\to\pi^0 e^+$ channel proton decay \cite{Hyper-Kamiokande:2018ofw}. 
 Super-K excludes the parameter space to the left of the purple solid line due to strong $p\to K^+\overline{\nu}$ channel decay. 
 To the left of the purple dashed line, the model can predict a proton lifetime that is within the sensitivity of JUNO \cite{JUNO:2015zny}.
 The narrow cyan region shows where the NANOGrav result can be explained at 95\% confidence \cite{NANOGrav:2023hfp}. } 
 \label{fig:ConandSen}
\end{figure}

The benchmark points can be located also in \figref{fig:ConandSen}, which presents the testability of this model in the $M_{\rm SUSY}-m_{\widetilde{W}}$ plane. The hatched region on the top-left indicates ($m_{\widetilde{W}} > M_{\rm SUSY}$)  is disfavoured in most SUSY scenarios. Gauge unification excludes the grid region on the bottom-left and the purple and orange solid lines are excluded by the current bounds on $\tau_{K^+\bar{\nu}}$ and $\tau_{\pi^0 e^+}$ from SUSY and non-SUSY contributions, respectively. The orange dashed line indicates the sensitivity of Hyper-K to $\pi^0 e^+$ channel decay, while the purple dashed line shows the potential target region of JUNO on $K^+\bar{\nu}$ channel decay.
A summary of the predictions of the model for different benchmark points is given in~\tabref{tab:BPs}.
BP1 has a phenomenology similar to non-SUSY SO(10) models \cite{Fu:2022lrn}.  
BP2 predicts proton decay rates that Hyper-K and JUNO can probe, although it has a string tension too high to be compatible with PTA observations. 
The region of the parameter space that predicts metastable strings that can explain the PTA observations, such as BP3, is in the region $G\mu \gtrsim 10^{-6}$ with $\sqrt{\kappa}\sim 8$ and can predict proton decay rates in the kaon channel at reach at JUNO, but interestingly, not at Hyper-K. 
Moreover, we notice that this region requires wino masses that are at reach at present or future colliders Refs.~\cite{ATLAS:2021moa,FCC:2018byv,CidVidal:2018eel,CEPCPhysicsStudyGroup:2022uwl}, with lower masses the lower the proton decay rate.
Interestingly, most of the parameter space that predicts proton decay rates too small to be observed by Hyper-K or JUNO is inconsistent with PTA observations. 

\begin{table}[h!]
\centering
\begin{tabular}{|c|c|c|c|}
\hline 
& HK sensitivity & JUNO target & NANOGrav15 \\\hline 
BP1 & testable & no signal & consistent \\ \hline 
BP2 & testable & targeted & inconsistent \\  \hline 
BP3 & no signal & targeted & support \\ \hline 
\end{tabular}
\label{tab:BPs}
\caption{ Complementary predictions of BP in the next-generation proton decay measurements and NANOGrav15.}
\end{table}

In the non-supersymmetric $SO(10)$ unification, it is also possible to obtain gauge unification, predict fermion masses and mixing and the dark matter \cite{Altarelli:2013aqa}. 
However, to achieve successful gauge unification and leptogenesis in the non-SUSY $SO(10)$ framework with a simple mass spectrum, the intermediate scale, that generates right-handed neutrinos masses and cosmic strings, is at least an order of magnitude smaller than the GUT scale 
\cite{King:2021gmj,Fu:2022lrn}. 
In that case, intermediate $U(1)$ symmetry breaking can only lead to stable cosmic strings and thus the resulting gravitational wave cannot provide an explanation of the spectrum observed by PTAs.
On the contrary, the intermediate $U(1)$ symmetry-breaking scale can be naturally close to the GUT scale in SUSY $SO(10)$ GUTs, which leads to the prediction of a metastable cosmic string network that can explain the PTA observations. 
Moreover, the proton decay predicted by SUSY GUTs is in general faster than that predicted by non-SUSY GUTs due to the additional kaon channel. 
As a result, the SUSY GUTs is more testable than the non-SUSY ones. 
As shown in \figref{fig:ConandSen}, the current bound on kaon channel proton decay rules out the possibility of low scale SUSY, which explains the absence of superpartners at LHC. 
Apart from the phenomenological perspectives, the hierarchy problem also motivates SUSY theoretically, further making SUSY $SO(10)$ an attractive candidate for unification.

\section{Summary}
We have presented a SUSY SO(10) GUT which can successfully predict fermion masses and mixing angles, leptogenesis, dark matter and can be tested via proton decay and GW signatures at next-generation experiments.
The $B-L$ breaking scale correlates with the SUSY breaking scale (the squark and slepton mass scale) via the gauge unification. 
The natural proximity of the GUT breaking scale and the $B-L$ breaking scale leads to metastable cosmic strings decaying to monopole-antimonopole pairs, and we find that the GW signal from metastable strings can be consistent with the NANOGrav 15-year data.  
Considering a split-SUSY scenario we found that proton decay measurements and PTA observations cover complementary regions of the parameter space. An eventual observation of proton decay from both the pion and kaon channels is not consistent with the current PTA observations.
Exploiting this complementarity, we can, therefore, test the majority of the parameter space.
Regarding the interpretation of the observed GW signal as generated by a metastable cosmic string network, we found that it is consistent with our model if the signal is partially inflated away, and that it is possible to fully test this possibility with the next-generation GW observatories and JUNO and/or collider searches.

\acknowledgements
We want to thank Stefan Antusch and Shaikh Saad for the helpful discussion. This work was partially supported by
Chinese Scholarship Council (CSC) Grant No. 201809210011 under agreements [2018]3101 and [2019]536, 
European Union's Horizon 2020 Research,
Innovation Programme under Marie Sklodowska-Curie grant agreement HIDDeN European ITN project (H2020-MSCA-ITN-2019//860881-HIDDeN),
National Natural Science Foundation of China (NSFC) under Grants No. 12205064,
Zhejiang Provincial Natural Science Foundation of China under Grant No. LDQ24A050002
STFC Consolidated Grant ST/T000775/1. 
This work used the DiRAC@Durham facility managed by the Institute for Computational Cosmology on behalf of the STFC DiRAC HPC Facility (www.dirac.ac.uk), which is part of the National e-Infrastructure and funded by BEIS capital funding via STFC capital grants ST/P002293/1, ST/R002371/1 and ST/S002502/1, Durham University and STFC operations grant ST/R000832/1. 
This work was partly performed in part at Aspen Center for Physics, which National Science Foundation supports grant PHY-2210452.

\appendix
\section{Gauge Unification \label{app:gauge_unification}}
A necessary condition for a realistic GUT model is that all gauge couplings unified to a single gauge coupling at a certain scale given by $g_{\rm GUT}$ and $M_{\rm GUT}$, respectively, up to matching conditions. We begin with the two-loop RGEs.
Given an interval of energy scale $Q \in (Q_0, Q_1)$, where the gauge symmetry is $G=H_1 \times H_2 \times \cdots \times H_n$ and no particles decouple in this period, the coupling $\alpha_i =g_i^2/(4\pi)$ for each gauge symmetry $H_i$ (for $i\in [1, \cdots, n]$) is described by the following differential equation
\bea
Q \frac{d\alpha_i}{d Q} = \beta_i (\alpha_i)\,.
\eea
The $\beta$ function, up to the two-loop level, is given by
\bea
\beta_i = - \frac{1}{2\pi} \alpha_i^2 ( b_i + \frac{1}{4\pi} \sum_{j} b_{ij} \alpha_j ) \,,
\eea
where $i\in [1, \cdots, n]$ for $H_n$, $g_i$ is the gauge coefficient of $H_i$, and $b_i$ and $b_{ij}$ refer to the normalised coefficients of one- and two-loop contributions, respectively. In the following, we neglect the Yukawa contribution to the R.G. running equations as it gives a subdominant contribution.
If the conditions $b_j \alpha_j(Q_0) \log(Q/Q_0) <1$ is satisfied, an analytical solution for these equations can be obtained \cite{Bertolini:2009qj}:
\bea
\alpha_i^{-1}(Q) &=& \alpha_i^{-1}(Q_0) - \frac{b_i}{2\pi}\log\frac{Q}{Q_0} \nonumber\\
&&+ \sum_j \frac{b_{ij}}{4\pi b_i} \log\left(1- \frac{b_j}{2\pi} \alpha_j(Q_0) \log \frac{Q}{Q_0}\right) \,.\quad\quad
\eea
In non-SUSY and SUSY, the coefficients $b_i$ and $b_{ij}$ are respectively given by 
\begin{widetext}
\begin{eqnarray}
\text{non-SUSY:} \quad b_i &=& - \frac{11}{3} C_2(H_i) + \frac{2}{3} \sum_\psi T(\psi_i) + \frac{1}{3} \sum_\phi T(\phi_i) \,, \nonumber\\
b_{ij} &=&
- \frac{34}{3} [C_2(H_i)]^2 \delta_{ij} + \sum_\psi T(\psi_i) [2 C_2(\psi_j) + \frac{10}{3} C_2(H_i) \delta_{ij}] + \sum_\phi T(\phi_i) [4 C_2(\phi_j) + \frac{2}{3} C_2(H_i) \delta_{ij}]\,;
\\
\;\text{SUSY:}\qquad b_i &=& - 3 C_2(H_i) + \sum_{\widetilde{\Phi}}T(\widetilde{\Phi}_i) \,, \nonumber\\
b_{ij} &=&
- 6 [C_2(H_i)]^2 \delta_{ij} + \sum_{\widetilde{\Phi}}2 T(\widetilde{\Phi}_i) [C_2(H_i) \delta_{ij} + 2C_2(\widetilde{\Phi}_j)] \,,
\eea
\end{widetext}
where $\phi$, $\psi$ and $\widetilde{\Phi}$ represent any complex scalar, chiral fermion and chiral superfield evolving in the scale between $Q_0$ and $Q_1$, respectively, $C_2(H_i)$ is the quadratic Casimir invariant of the adjoint presentation in the group $H_i$, $C_2(\phi_i)$ ($C_2(\widetilde{\Phi}_i)$) is quadratic Casimir invariant of the representation of the field $\phi$ (superfield $\widetilde{\Phi}$) in the group $H_i$, $T(\phi_i)$ ($T(\widetilde{\Phi}_i)$) is the Dynkin index of the field $\phi$ (superfield $\widetilde{\Phi}$) in group $H_i$, and the summation goes over all fields (superfields). In our model, values of coefficients $b_i$ and $b_{ij}$ in each interval of the energy scale from $M_{\rm GUT}$ down to $M_Z$ are shown in Table~\ref{tab:beta}. 

At the intermediate scale, where a larger symmetry is broken to its sub-symmetry, gauge couplings between them satisfy matching conditions. Here we list one-loop matching conditions that appear in the GUT-breaking chains. For a simple Lie group $H_{i+1}$ broken to subgroup $H_i$ at the scale $Q = M_I$, the one-loop matching condition in the $\overline{\rm MS}$ scheme is given by \cite{Chakrabortty:2017mgi}
\bea
&&H_{i+1} \to H_i\,: \nonumber\\
&&\alpha_{H_{i+1}}^{-1}(M_I) - \frac{1}{12\pi} C_2(H_{i+1}) = \alpha_{H_i}^{-1}(M_I) - \frac{1}{12\pi} C_2(H_i) \,.\nonumber\\
\eea
Above the SUSY scale, the couplings must run in the $\overline{\rm D.R.}$ scheme to preserve the supersymmetry \cite{Martin:1993yx}. The relation of couplings in the $\overline{\rm MS}$ scheme and $\overline{\rm DR}$ scheme is described by
\bea
\alpha_{\rm DR}^{-1} &=& \alpha_{\rm MS}^{-1} - \frac{1}{12\pi} C_2(H_i) \,.
\eea
Thus the one-loop matching condition in the $\overline{\rm DR}$ scheme is simply
\bea
H_{i+1} \to H_i\,: \quad
\alpha_{H_{i+1}}^{-1}(M_I)&=& \alpha_{H_i}^{-1}(M_I) \,.
\eea
For $G_{\rm int}\to G_{\rm MSSM}$, we encounter the breaking, $SU(2)_R \times U(1)_{X} \to U(1)_Y$, where $U(1)_X$ is identical to $U(1)_{B-L}$ with the charge normalised as $X = \sqrt{3/8} (B-L)$. The matching condition in $\overline{\rm DR}$ scheme is given by
\bea
&&SU(2)_R \times U(1)_X \to U(1)_Y \,: \nonumber\\
&&\frac{3}{5} \alpha_{2R}^{-1}(M_{B-L}) + \frac{2}{5}\alpha_{1X}^{-1}(M_{B-L}) = \alpha_{1Y}^{-1}(M_{B-L}) \,. \quad\quad\quad
\eea
Applying the matching conditions, all gauge couplings of the subgroups unify into a single gauge coupling of $SO(10)$ at the GUT scale, $M_{\rm GUT}$. 
This condition restricts both the GUT and intermediate scales for each breaking chain. 
We denote the mass of the heavy gauge boson masses associated with $SO(10)$ breaking as $M_{\rm GUT}$ while $M_{B-L}$ and $M_{\rm SUSY}$ are associated to the breaking of $G_{\rm int}$ and $G_{\rm MSSM}$, respectively. 
\begin{table*}[!] 
\begin{center}
\begin{tabular}{| c l |}
\hline \hline 
$SO(10)$ & broken at $Q=M_{\rm GUT}$ \\\hline &\\[-3mm]
$\Bigg\downarrow$ & 
$\{b_i\} = \begin{pmatrix}-3 \\ 2 \\ -4 \\ \frac{21}{2} \end{pmatrix} \,, \quad
\{b_{ij} \} = \begin{pmatrix}
 14 & 9 & 9 & 1 \\
 24 & 32 & 6 & 3 \\
 24 & 6 & 56 & 15 \\
 8 & 9 & 45 & 34 \\
 \end{pmatrix}$ \\ & \\[-3mm]\hline
$G_{\rm int}$ & broken at $Q=M_{B-L}$ \\\hline
& \\[-3mm]
$\Bigg\downarrow$ & 
$\{b_i\} = \begin{pmatrix} -3 \\ 1 \\ \frac{33}{5} \end{pmatrix} \,,\quad
\{b_{ij} \} = \begin{pmatrix} 
 14 & 9 & \frac{11}{5} \\
 24 & 25 & \frac{9}{5} \\
 \frac{88}{5} & \frac{27}{5} & \frac{199}{25} \\
\end{pmatrix}$ \\[-3mm] & \\\hline
$G_{\rm MSSM}$ & broken at $Q=M_{\rm MSSM}$ \\\hline
& \\[-3mm]
$\Bigg\downarrow$ & 
$\{b_i\} = \begin{pmatrix} -5 \\ -\frac{7}{6} \\ \frac{9}{2} \end{pmatrix} \,, \quad
\{b_{ij} \} = \begin{pmatrix}
22 & \frac{9}{2} & \frac{11}{10} \\
12 & \frac{106}{3} & \frac{6}{5} \\
\frac{44}{5} & \frac{18}{5} & \frac{104}{25} 
\end{pmatrix}$ \\[-3mm] &
\\\hline
\hspace{2mm}wino and gluino decoupling \hspace{2mm} & happening at $Q=M_{\widetilde{W}}$ \\\hline
& \\[-3mm]
$\Bigg\downarrow$ & 
$\{b_i\} = \begin{pmatrix} -7 \\ -\frac{5}{2} \\ \frac{9}{2} \end{pmatrix} \,,\quad
\{b_{ij} \} = \begin{pmatrix} 
-26 & \frac{9}{2} & \frac{11}{10} \\
12 & 14 & \frac{6}{5} \\
\frac{44}{5} & \frac{18}{5} & \frac{104}{25}
\end{pmatrix}$ 
\\[-3mm] & \\\hline
\hspace{2mm}bino and higgsino decoupling \hspace{2mm} & happening at $Q=M_{\widetilde{B}}$ \\\hline
& \\[-3mm]
$\Bigg\downarrow$ & 
$\{b_i\} = \begin{pmatrix} -7 \\ -\frac{19}{6} \\ \frac{41}{10} \end{pmatrix} \,,\quad
\{b_{ij} \} = \begin{pmatrix} 
-26 & \frac{9}{2} & \frac{11}{10} \\
12 & \frac{35}{6} & \frac{9}{10} \\
\frac{44}{5} & \frac{17}{10} & \frac{199}{50} 
\end{pmatrix}$ 
\\[-3mm] & \\\hline
$G_{\rm SM}$ & \\
\hline \hline
\end{tabular}
\end{center}
\caption{Coefficients $b_i$ and $b_{ij}$ of gauge coupling $\beta$ functions appearing in the specified breaking chain. In this table, we identify the scale of gauge symmetry as the corresponding heavy gauge boson mass from the EFT point of view. The SUSY breaking scale $M_{\rm MSSM}$ is regarded as the unified mass of all sfermions. In the split SUSY, gauge bosons and Higgs superpartners are allowed to be different from and usually a few orders of magnitude lower than the SUSY-breaking scale. We also consider the possibility of having a further gap between the wino and gluino masses and the bino and higgsino masses. We include their threshold effect in the RG running. \label{tab:beta}}.
\end{table*}

\section{Matter and Higgs Decomposition}\label{sec:mattercontent}

The ${\bf 45}$ and ${\overline{\bf 126}}$ 
are required for gauge symmetry breaking, and we list their decomposition under $G_{\rm int} \equiv SU (3)_c \times SU (2)_L \times SU (2)_R \times U(1)_{B-L} \times \text{SUSY} $ and 
$G_{\rm SM} \equiv SU(3)_c \times SU(2)_L \times U(1)_Y$ in \tabref{tab:decomposition_H_breaking}.

\begin{table}[!] 
\begin{center}
\begin{tabular}{| c | c | c |}
\hline \hline 
$SO(10)$ \quad & 
${\bf 45}$ &
${\overline{\bf 126}}$
\\\hline
$G_{\rm int}$ &
$({\bf 1}, {\bf 1},{\bf 1}, 0)$ & 
$({\bf 1}, {\bf 1}, {\bf 3}, -1)$ \\\hline
$G_{\rm SM}$ & 
({\bf 1}, {\bf 1}, 0) &
$({\bf 1}, {\bf 1}, 0)_S$ \\
\hline \hline
\end{tabular}
\end{center}
\caption{{Decomposition of the Higgses which induce spontaneous symmetry breaking at each step of the breaking chain. Each Higgs (from left to right) is eventually decomposed to a singlet whose non-vanishing VEV preserves the symmetry $G_I$ (for $I = 3,2,1,{\rm SM}$) in the same row but breaks larger symmetries. The subscript $S$ refers to ${\overline{\bf 126}}$ containing a SM singlet scalar whose VEV generates the RHN masses via B-L symmetry breaking.} \label{tab:decomposition_H_breaking}}
\end{table}
We include two further Higgs multiplets, ${\bf 10}$ and ${\bf 120}$, to generate the Standard Model fermion masses \footnote{It has been argued in Ref.~\cite{Babu:2018tfi} that a ${\bf 10}$ and a ${\bf 126}$ are enough to reproduce the fermion data.}. 
We show their decompositions under $G_{1}$ and $G_{\rm SM}$ in \tabref{tab:decomposition_H_mass} where the subscript is used to distinguish fields with the same representation.
In \tabref{tab:decomposition_H_mass}, $\overline{\bf 126}$ is the same Higgs used in the breaking $G_{\rm int} \to G_{\rm SM}$ which has the dual role of spontaneous symmetry breaking and flavour structure generation.
\begin{table*}[!] 
\begin{center}
\begin{tabular}{| c | c | c | c |}
\hline \hline 
$SO(10)$ & ${\bf 10}$ & ${\overline{\bf 126}}$ & ${\bf 120}$ 
\\\hline
\multirow{2}*{$G_{\rm int}$} & 
$({\bf 1}, {\bf 2},{\bf 2}, 0)_1$ &
$({\bf 1}, {\bf 2},{\bf 2},0)_2$ &
$({\bf 1}, {\bf 2},{\bf 2}, 0)_{3,4}$ \\
&& $+ ({\bf 1}, {\bf 1}, {\bf 3}, -1)$ & \\\hline
\multirow{3}*{$G_{\rm SM}$} & $({\bf 1}, {\bf 2}, - 1/2)_{h_{\bf 10}^{u}}$ &
$({\bf 1}, {\bf 2}, - 1/2)_{h_{\overline{\bf 126}}^{u}}$ &
$({\bf 1}, {\bf 2}, - 1/2)_{h_{\bf 120}^{u}, {h_{\bf 120}^{u'}}}$ \\
&
$+ ({\bf 1}, {\bf 2}, + 1/2)_{h_{\bf 10}^{d}}$ &
$+ ({\bf 1}, {\bf 2}, + 1/2)_{h_{\overline{\bf 126}}^{d}}$ &
$+ ({\bf 1}, {\bf 2}, + 1/2)_{h_{\bf 120}^{d}, {h_{\bf 120}^{d'}}}$ \\
&& $+ ({\bf 1}, {\bf 1},0)_S$ & \\
\hline \hline
\end{tabular}
\end{center}
\caption{Decomposition of Higgses responsible for the fermion mass generation. $\overline{\bf 126}$ is the same Higgs as shown in Table~\ref{tab:decomposition_H_breaking} and it is responsible for both the breaking $G_{\rm int} \to G_{\rm SM}$ and right-handed neutrino mass generation. $({\bf 1}, {\bf 1},0)_S$ is the same singlet given in Table~\ref{tab:decomposition_H_breaking}. }\label{tab:decomposition_H_mass}
\end{table*}

\begin{table}[!] 
\begin{center}
\begin{tabular}{| c | c |}
\hline \hline 
$SO(10)$ & ${\bf 16}$
\\\hline & \\[-5mm]
\multirow{2}*{$G_{\rm int}$} &
$({\bf 3}, {\bf 2},{\bf 1}, 1/6)_{Q_L} + (\overline{\bf 3}, {\bf 1},{\bf 2}, -1/6)_{Q_R^c}$ \\ 
& $+ ({\bf 1}, {\bf 2}, {\bf 1}, -1/2)_{l_L} + ({\bf 1}, {\bf 1},{\bf 2}, 1/2)_{l_R^c}$ \\[1mm]\hline
& \\[-5mm]
\multirow{2}*{$G_{\rm SM}$} &
$({\bf 3}, {\bf 2},1/6)_{Q_L} + (\overline{\bf 3}, {\bf 1},-2/3)_{u_R^c} + (\overline{\bf 3}, {\bf 1},1/3)_{d_R^c}$ \\
& $+({\bf 1}, {\bf 2}, -1/2)_{l_L}\!+\!({\bf 1}, {\bf 1}, 0)_{\nu_R^c} + ({\bf 1}, {\bf 1},1)_{e_R^c}$
\\[1mm]
\hline \hline
\end{tabular}
\end{center}
\caption{{Decomposition of the matter multiplet ${\bf 16}$ in each step of the breaking chain.} \label{tab:decomposition_f}}
\end{table}

The fermions are arranged as a ${\bf 16}$ of $SO(10)$ and follow the decomposition given in \tabref{tab:decomposition_f} where $L$ ($R$) denote the left-handed (right-handed) fermions of $G_3$ which contains the SM left-handed (right-handed) fermions where $Q_{L (R)}$ and $\ell_{L (R)}$ are the quark and leptonic $SU(2)_{L(R)}$ doublets, respectively, and $u_{R}, d_{R}, e_{R}$, and $\nu_{R}$ are the quark and lepton $SU(2)_{L}$ singlets, respectively.

\section{Correlations of fermion masses and mixing}\label{sec:matter}
In this section, we present the correlations of masses and mixing between quarks and leptons and predict the RHN masses using the model we discussed in the previous section. 
We parametrise the up, down, neutrino, charged lepton Yukawa couplings and right-handed neutrino mass matrix following the left-right notation in Refs.~\cite{Altarelli:2010at,Dutta:2004zh,Dutta:2005ni,Dutta:2009ij}, respectively, as follows: 
\bea
&& Y_u = h + r_2 f + i\, r_3 h^\prime \,, \quad
Y_d = r_1 (h+ f + i\, h^\prime) \,, \nonumber\\
&&Y_\nu = h - 3 r_2 f + i\, c_\nu h^\prime\,, \quad
Y_e = r_1 (h - 3 f + i\, c_e h^\prime) \,,  \nonumber\\
&& M_{\nu_R} = f \, \frac{\sqrt{3} \,r_1}{U_{12}} v_S\,,
\label{eq:Yukawa_Hermitian}
\eea
where 
\bea
&& h = Y_{\bf 10} V_{11} \,, \;
f = Y_{\overline{\bf 126}} \frac{U_{12}}{\sqrt{3}} \frac{V_{11}}{U_{11}} \,, \nonumber\\
&& h^\prime= -i\, Y_{\bf 120} \left(U_{13}+ U_{14}/\sqrt{3}\right) \frac{V_{11}}{U_{11}}\,, \nonumber\\
&& r_1 = \frac{U_{11}}{V_{11}} \,, \quad
r_2 = \frac{V_{12}}{U_{12}} \frac{U_{11}}{V_{11}}\,,\quad
r_3 = \frac{V_{13}+ V_{14}/\sqrt{3}}{U_{13}+ U_{14}/\sqrt{3}} \frac{U_{11}}{V_{11}}\,,\nonumber\\
&& c_e= \frac{U_{13}- \sqrt{3} U_{14}}{U_{13}+ U_{14}/\sqrt{3}}\,, \quad
c_\nu= \frac{V_{13}- \sqrt{3} V_{14}}{U_{13}+ U_{14}/\sqrt{3}} \frac{U_{11}}{V_{11}}\,,
 \label{eq:parametrise_Yukawa}
\eea
and $V_{ij}$ and $U_{ij}$ denotes the mixing between the mass and interaction basis of Higgs doublets in the up and down sectors, respectively, before the SUSY breaking \cite{Dutta:2004zh,Dutta:2005ni}, and $v_S$ is the VEV of singlet component of $\overline{\bf 126}$ that gives mass to the RHNs. The light neutrino mass matrix, $M_\nu$, is obtained by
\bea
M_\nu &=&
m_0 Y_\nu f^{-1} Y_\nu\,,
\eea 
where $m_0= - \frac{U_{12}}{\sqrt{3} \,r_1} \frac{v_{u}^2}{v_S}$. Once the scale $v_S$ is determined, $U_{12}$ can be solved as 
\bea
U_{12}=- \sqrt{3} \,r_1 \frac{m_0 v_S}{v_{u}^2}\,.\label{eq:U12}
\eea 
The most general form of Yukawa couplings and neutrino mass matrix includes many free parameters.
A considerable reduction in the number of parameters can be achieved by considering only
the Hermitian case for all fermion Yukawa couplings matrices $Y_u$, $Y_d$, $Y_\nu$ and $Y_e$ (and $M_R$ should be real as a consequence of the Majorana nature for right-handed neutrinos).
Such a reduction can result from spontaneous CP violation \cite{Grimus:2006bb,Grimus:2006rk}, which assumes that there exists a CP symmetry above the GUT scale, leading to real-valued $Y_{\bf 10}$, $Y_{\overline{\bf 126}}$ and $Y_{\bf 120}$, and the CP is broken by some complex VEVs of Higgs multiplets during GUT or intermediate symmetry breaking.
As a result, $h$, $f$ and $h'$, as well as all parameters on the right-hand side of Eq.~\eqref{eq:Yukawa_Hermitian}, are real. Since $h'$ is antisymmetric, we arrive at Hermitian Dirac Yukawa coupling matrices $Y_u$, $Y_d$, $Y_\nu$ and $Y_e$. The Higgs mixing elements $V_{13}$, $V_{14}$ and $U_{13}$, $U_{14}$ are also purely imaginary with the relations 
\bea
\frac{V_{13}}{V_{14}} = \frac{c_\nu/r_3+3}{\sqrt3(c_\nu/r_3-1)}\,, \quad
\frac{U_{13}}{U_{14}} = \frac{c_e+3}{\sqrt3(c_e-1)}\,. \label{eq:relation_1}
\eea
This texture has been widely applied in the literature, e.g., Refs.~\cite{Dutta:2004hp,Dutta:2004zh,Joshipura:2011nn}. The resulting fermion mass matrices conserve parity symmetry $L \leftrightarrow R$ \cite{Grimus:2006rk} and following from the assumption that there is no C.P. violation in the Higgs sector, apart from that of ${\bf 120}$, $r_1$, $r_2$, $r_3$, $c_e$, and $c_\nu$ are all real parameters
resulting in a real symmetric right-handed neutrino mass matrix, $M_{\nu_R}$. 
The CP symmetry in the Yukawa coupling is spontaneously broken after the Higgses gain VEVs. 

\subsection{Procedure to fit the quark and lepton flavour data}

For simplicity, we assume that $r_3 = 0$, which implies that the imaginary part of $Y_u$ vanishes. As a result, the relation between $V_{13}$ and $V_{14}$ in \eqref{eq:relation_1} is no longer valid. Instead, there is a simpler relation that reads $V_{14}=-\sqrt3 V_{13}$. It is convenient to write the up-type Yukawa in the diagonal basis
\bea
Y_u = h + r_2 f = {\rm diag}\{ \eta_u y_u, \eta_c y_c, \eta_t y_t\}\,,
\eea
which can be achieved via a real-orthogonal transformation on the fermion flavours without changing the Hermitian property of $Y_d$, $Y_e$, and $Y_\nu$.
In the above, $\eta_{u,c,t} = \pm 1$ refer to signs that the real-orthogonal transformation cannot determine. While $\eta_t = +1$ can be fixed by making an overall sign rotation for all Yukawa matrices, the remaining signs, $\eta_u$ and $\eta_c$, cannot be fixed and are randomly varied throughout our analysis.
In the basis of the diagonal up-quark mass matrix, $Y_d$ is given by
\bea
Y_d = P_a V_{\rm CKM} \, {\rm diag}\{ \eta_d y_d, \eta_s y_s, \eta_b y_b\} \,V_{\rm CKM}^\dag P_a^* \,,
\eea 
where again $\eta_{d,s,b} = \pm 1$ represent the signs of eigenvalues, and $V_{\rm CKM}$ is the CKM matrix parametrised in the following form
\begin{widetext}
\bea
V_{\rm CKM} = \left(
\begin{array}{ccc}
 c_{12} c_{13} & s_{12} c_{13} & s_{13} e^{-i \delta_q } \\
 - s_{12} c_{23}- c_{12} s_{13} s_{23} e^{i \delta_q } & c_{12} c_{23}- s_{12} s_{13} s_{23} e^{i \delta_q } & c_{13} s_{23} \\
 s_{12} s_{23}- c_{12} s_{13} c_{23} e^{i \delta_q } & - c_{12} s_{23}- s_{12} s_{13} c_{23} e^{i \delta_q } & c_{13} c_{23} \\
\end{array}
\right) \,,
\eea 
where $s_{ij} = \sin \theta_{ij}^q$, $c_{ij} = \cos \theta_{ij}^q$ and 
$P_a= {\rm diag}\{ e^{ia_1}, e^{ia_2}, 1\}$. The matrices $h$, $f$ and $h^\prime$ are then expressed in terms of $Y_u$ and $Y_d$
\bea
&& h =- \frac{Y_u}{r_2-1} + \frac{r_2 {\rm Re} Y_d}{r_1(r_2-1)} \,, \quad 
f =\frac{Y_u}{r_2-1}- \frac{{\rm Re} Y_d}{r_1(r_2-1)} \,, \nonumber\\
&& h^\prime = i \, \frac{{\rm Im} Y_d}{r_1}\,,
\eea
where $Y_\nu$, $Y_e$ are 
\bea
Y_\nu &=& - \frac{3r_2+1}{r_2-1} Y_u + \frac{4r_2}{r_1(r_2-1)}{\rm Re} Y_d + i \frac{c_\nu}{r_1} {\rm Im} Y_d \,, \nonumber\\
Y_e &=& - \frac{4r_1}{r_2-1} Y_u + \frac{r_2+3}{r_2-1}{\rm Re} Y_d +i c_e {\rm Im} Y_d\,. \label{eq:Yukawa_lepton}
\eea 
The light neutrino mass matrix can be expressed as
\bea \label{eq:M_nu}
M_\nu &=& m_0 \left( \frac{8r_2(r_2+1)}{r_2-1}Y_u - \frac{16 r_2^2}{r_1(r_2-1)} {\rm Re} Y_d + \frac{r_2-1}{r_1} \left(r_1Y_u + i c_\nu {\rm Im} Y_d \right) \left(r_1Y_u - {\rm Re} Y_d \right)^{-1} \left(r_1Y_u - i c_\nu {\rm Im} Y_d \right) \right)\,.\quad\quad
\eea 
\end{widetext}
Using this parametrisation, all six quark masses and four CKM mixing parameters are treated as inputs, and we are then left with seven parameters ($a_1$, $a_2$, $r_1$, $r_2$, $c_e$, $c_\nu$, and $m_0$) to fit eight observables, including three Yukawa couplings $y_e$, $y_\mu$, $y_\tau$, two neutrino mass-squared differences $\Delta m^2_{21}$, $\Delta m^2_{31}$ and three mixing angles $\theta_{12}$, $\theta_{13}$, $\theta_{23}$, where the leptonic CP-violating phase, $\delta$, will be treated as a prediction.

By fitting the fermion mass and mixing, the matrices $h$, $f$, $h'$ and parameters $a_1$, $a_2$, $r_1$, $r_2$, $c_e$, $c_\nu$, and $m_0$ can be fully determined. To perform the parameter scan, and find viable regions of the model parameter space that postdict the quark and predict the leptonic data, we run all the SM Yukawa couplings to the GUT scale (using two-loop RGEs, appropriate matching between scales and threshold corrections) using {\sc REAP}\cite{Antusch:2005gp,Antusch:2015nwi} and {\sc SARAH} \cite{Staub:2015kfa}. We then scan the free parameters of the GUT model as described above and assess how well they fit the leptonic data using the statistical measure below:
\begin{figure*}[t!]
 \centering
 \includegraphics[width = 0.9\textwidth]{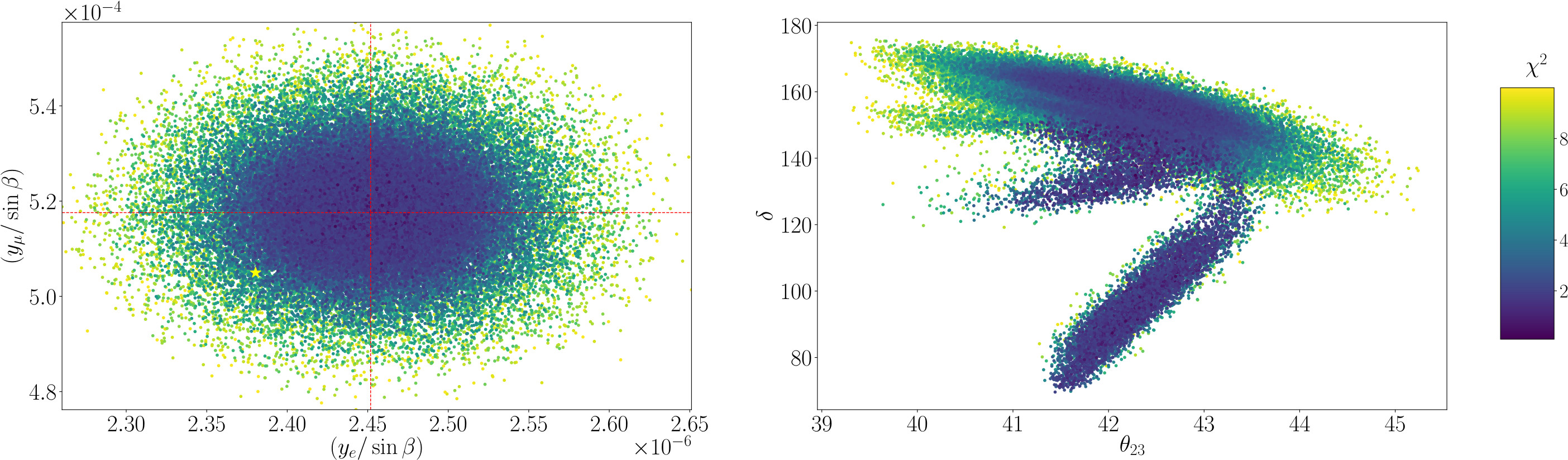}
 \caption{\label{fig:scanpred}All coloured points show regions of the parameter space which fit the lepton sector with $\chi^2 < 10$. The yellow star shows a benchmark point in the scan with $\eta_B \sim 6.16\times 10^{-10}$. All the points in the scan predict $-11\lesssim \log_{10}(\eta_B)\lesssim-8$.}
\end{figure*}
\bea 
\chi^2 = \sum_n \left[ \frac{{\cal O}_n({\cal P}_m)-{\cal O}_n^{\rm bf}}{\sigma_{{\cal O}_n}} \right]^2 \,,
\eea 
where ${\cal P}_m \in \{a_1, a_2, r_1, r_2, c_e, c_\nu, m_0, \eta_q\}$ and ${\cal O}_n \in \{m_e, m_\mu, m_\tau, \theta_{12}, \theta_{13}, \theta_{23}, \Delta m^2_{21}, \Delta m^2_{31}\}$.

Then Yukawa couplings for $SO(10)$ $\bf 16$ multiplets can be expressed as 
\bea
Y_{\bf 10} = \frac{h}{V_{11}} \,, \; 
Y_{\overline{\bf 126}} &=& - f\, \frac{v_{u}^2}{m_0\, v_S}\,, \; 
Y_{\bf 120}= i\, h^\prime\, \frac{c_\nu}{4 V_{13}},\quad\quad
\label{eq:SO10_Yukawas}
\eea
this will be relevant for the subsequent discussion on proton decay, tightly linked with the scan of fermion masses and mixing since it is mediated by the Higgs colour triplet and the operators computed from the same superpotential.
Apart from the matrices and parameters that can be fixed by fermion mass and mixing, there are three parameters: two Higgs mixing elements $V_{11}$ and $V_{13}$ and $\tan\beta$. 
Apart from the equations above, there are also 
\bea
&&U_{11}=r_1 V_{11}\,, \quad
V_{12}=\frac{r_2}{r_1} U_{12}\,,\quad
U_{13}=\frac{2r_1}{c_\nu}\frac{c_e+3}{c_e+1} V_{13} \,,\nonumber\\
&&U_{14}=\frac{2\sqrt3 r_1}{c_\nu} \frac{c_e-1}{c_e+1} V_{13}\,,\quad
V_{14}=-\sqrt3 V_{13}\,,\quad
\eea
while $U_{12}$ can be solved using \eqref{eq:U12}.
Each element in the mixing matrices has to satisfy the unitarity.
We show a subset of our predictions from the scan in \figref{fig:scanpred}. In the left (right) panel, we show the predictions for the muon ($\delta$ phase) Yukawa versus the electron Yukawa ($\theta_{23}$). We note that the predictions are consistent with the experimental values at the 3$\sigma$ level. While all values of $\delta$ can be accommodated, the model prefers the atmospheric mixing angle to be in the lower octant. Moreover, the model strongly prefers normally ordered light neutrino masses.

\subsection{A benchmark study}\label{sec:BM}
We considered a benchmark point of our scan (yellow star), achieving successful leptogenesis, giving $\eta_B \sim 6.16\times 10^{-10}$. 
Inputs and predictions of fermion Yukawas and mixing parameters are shown in \tabref{tab:benchmark}.
From this, the Yukawa and neutrino mass matrices are obtained: 
\onecolumngrid
\begin{table*}[!] 
\begin{center}
\begin{tabular}{ l |c c c c c c}
\hline \hline 
Inputs & $a_1$ & $a_2$ & $c_\nu$ & $m_0$ & \hspace{-2mm}$(\eta_u, \eta_c, \eta_t; \eta_d, \eta_s, \eta_b)$\\ 
 & $35,40^\circ$ & $221.27^\circ$ & -1.49 & 44.24\,meV & \hspace{-2mm}$(-,+,+; +,-,-)$ \\\hline
Outputs & $\theta_{13}$ & $\theta_{12}$ & $\theta_{23}$ & $\delta$ & $m_1$ \\
 & $8.66^\circ$ & $33.19^\circ$ & $44.14^\circ$ & $131.57^\circ$ & $5.29$ meV \\\cline{2-6}
 ($\chi^2=8.22$)& $m_{\beta\beta}$ & & $M_{N_1}$ & $M_{N_2}$ & $M_{N_3}$ & \\
& 5.76\,meV & & \!$8.18 \cdot 10^{11}$\,GeV\! & \!$1.53 \cdot 10^{12}$\,GeV\! & \!$4.67 \cdot 10^{13}$\,GeV\! \\\cline{2-6}
\hline \hline
\end{tabular}
\end{center}
\caption{Inputs and predictions of neutrino masses and the benchmark point mixing parameters fully satisfy all experimental data. Charged fermion masses and CKM mixing are all fixed at experimental best-fit values. Neutrino masses with normal ordering are predicted. 
\label{tab:benchmark}}
\end{table*}
\twocolumngrid
\begin{widetext}
\be
\label{eq:benchmark1}
Y_u/\cos\beta = \left(
\begin{array}{ccc}
 3.04\cdot 10^{-6} & 0 & 0 \\
 0 & 0.00149 & 0 \\
 0 & 0 & 0.489 \\
\end{array}
\right)\,,
\ee
\be
Y_d/\sin{\beta} = 10^{-2} \cdot \left(
\begin{array}{ccc}
-0.0051423 & -0.0039347 + 0.00050024i & 0.0019310 - 0.0013028i \\
-0.0039347 - 0.00050024i & 0.12261 & -0.19878 - 0.17455i \\
0.0019310 + 0.0013028i & -0.19878 + 0.17455i & 0.61483 \\
\end{array}
\right)\,,
\ee
\be
Y_e/\sin{\beta} = 10^{-2} \cdot \left(
\begin{array}{ccc}
-0.0011843 & -0.00090268i + 0.0064489i & 0.00044301 - 0.016795i \\
-0.00090268i - 0.0064489i & 0.0050731 & -0.0045604 - 0.22502i \\
0.00044301i + 0.016795i & -0.0045604 + 0.22502i & 0.88287 \\
\end{array}
\right)\,,
\ee
\be
M_\nu = \cdot \left(
\begin{array}{ccc}
6.259 + 6.680i & 4.786 + 2.162i & -2.167 + 1.916i \\
4.786 + 2.162i & -13.836 + 0.1096i & 26.166 + 8.005i \\
-2.167 + 1.916i & 26.166 + 8.005i & -15.644 - 25.224i \\
\end{array}
\right)~{\rm meV} \,.
\ee
\end{widetext}
Diagonalisation of $Y_e$ and $M_\nu$ gives rise to the lepton masses and mixing, and the above benchmark provides 
 $\chi^2 =8.22$. \\

\section{Proton decay} \label{app:proton_decay}

Unified theories may contain gauge or scalar bosons that mediate baryon (B) number and lepton number (L) violating processes and hence can cause the decay of nucleons. 
The dominant mediator is the heavy colour gauge boson which leads to proton decay where the most constrained channel is $p\to \pi^0 e^+$ with the lifetime $\tau_{\pi^0 e^+} \gtrsim 2.4 \times 10^{34}$~years set by Super-K \cite{Super-Kamiokande:2020wjk}. 
Due to the direct correlation $\tau_{\pi^0 e^+} \propto M_{\rm GUT}^4$, this channel usually provides the most constraining limit on the GUT scale, $M_{\rm GUT} \gtrsim 3 \times 10^{15}$~GeV, almost regardless of the breaking chains of $SO(10)$ \cite{King:2021gmj}. 

The following section discusses proton decay induced by GUT and SUSY symmetry breaking, respectively.

\subsubsection{Pion channel}
The computation of the proton lifetime in this channel is identical in both the SUSY and the non-SUSY case, and it is carried by the heavy SO(10) gauge bosons. The relevant dimension-six operators are written as
\begin{eqnarray} \label{eq:D6}
\frac{\epsilon_{\alpha\beta}}{\Lambda^2} &&\left[
(\overline{u_R^{c}} \gamma^\mu Q_\alpha)(\overline{d_R^{c}} \gamma_\mu L_\beta) +
(\overline{u_R^{c}} \gamma^\mu Q_\alpha)(\overline{e_R^{c}} \gamma_\mu Q_\beta) \right. \nonumber\\&&\left.
+
(\overline{d_R^{c}} \gamma^\mu Q_\alpha)(\overline{u_R^{c}} \gamma_\mu L_\beta) +
(\overline{d_R^{c}} \gamma^\mu Q_\alpha)(\overline{\nu_R^{c}} \gamma_\mu Q_\beta) \right ], \nonumber\\
\end{eqnarray}
where colour and flavour indices have been suppressed and 
 $\Lambda \simeq \sqrt{2} M_{\rm GUT}/g_{\rm GUT}$ denotes the UV completion scale. 
The proton lifetime is \cite{King:2021gmj}: 
\be
\ba
&\Gamma\left(p \rightarrow \pi^0+e^{+}\right)\\=&\frac{m_p}{32 \pi}\left(1-\frac{m_{\pi^0}^2}{m_p^2}\right)^2 A_L^2 \times \\
& {\left[A_{S L} \Lambda_1^{-2}\left(1+\left|V_{u d}\right|^2\right)\left|\left\langle\pi^0\left|(u d){R u_L}\right| p\right\rangle\right|^2\right.} \\ & 
\left.+A_{S R}\left(\Lambda_1^{-2}+\left|V_{u d}\right|^2 \Lambda_2^{-2}\right)\left|\left\langle\pi^0\left|(u d)_L u_L\right| p\right\rangle\right|^2\right]\,,
\ea
\ee
The enhancement factors denoted as $A_L$, $A_{S.L.}$, and $A_{S.R.}$, correspond to the influence of long and short-range effects on proton decay. The relevant hadronic matrix element for this particular decay mode is $\langle \pi^0 | (ud)_{L,R} u_L | p \rangle$, which has been determined through a QCD lattice simulation \cite{Aoki:2017puj}.
The long-range effect incorporates the renormalisation enhancement from the proton decay process (at scales $\sim 1$ GeV) to the electroweak scale (defined as the mass of the Z boson at the scale of $M_Z$). This enhancement factor, calculated at the two-loop level, for is $A_L = 1.247$ in the studies \cite{Ellis:2020qad}.
In the SUSY case, we used $A_L = 0.4$ \cite{Ellis:1981tv,Hisano:1992jj}. The short-range factors are determined through the renormalisation group equations, which span from the scale $M_Z$ to $M_{\rm GUT}$:
\begin{align}
  \begin{aligned}
  A_{S L(R)}=\prod_A^{M_Z \leqslant M_A \leqslant M_{\rm GUT}} \prod_i\left[\frac{\alpha_i\left(M_{A+1}\right)}{\alpha_i\left(M_A\right)}\right]^{\frac{\gamma_{i L(R)}}{b_i}}\,,
  \end{aligned}
\end{align}
where $\gamma_i$ are the anomalous dimensions and $b_i$ the one-loop $\beta$ coefficients. 

\subsection{Kaon channel}
For the computation of the proton decay in the kaon channel, we followed the treatment of Ref.~\cite{Goh:2003nv,Severson:2015dta}. The baryon number violating interaction is mediated by the Higgs colour triplet with mass $M_T \simeq M_{\rm GUT}$. 
The terms in the superpotential, which we consider, are the same as the Yukawa sector in Eq.~(2). Let us consider the effective superpotential from which we can infer all the 5-dimensional operators contributing to the kaon channel, 
\be
\ba
W_{\Delta B=1}=\frac{\epsilon_{a b c}}{M_T}&\left(C_{\alpha \beta \gamma \delta}^L Q_\alpha^a Q_\beta^b Q_\gamma^c L_l\right.\nonumber\\
&\left.+C_{[\alpha \beta \gamma] \delta}^R U_\alpha^{\mathcal{C} a} D_\beta^{\mathcal{C} b} U_\gamma^{\mathcal{C} c} E_\delta^{\mathcal{C}}\right)\,,
\label{eq:effective_superp3}
\ea
\ee 
where the matter should be understood as chiral superfield, $a,b,c$ are colors indices and $\alpha,\beta,\gamma,\delta$ are flavour indexes and $C_{ijkl}^L$ and $C_{ijkl}^R$ are the Wilson coefficients, and in the flavour basis they are expressed as
\be
\begin{aligned}
C_{\alpha \beta \gamma \delta}^R & =(Y_{\mathbf{10}})_{\alpha \beta} (Y_{\mathbf{10}})_{\gamma \delta}+x_1 (Y_{\mathbf{\overline{126}}})_{\alpha \beta} (Y_{\mathbf{\overline{126}}})_{\gamma \delta}\\ &
+x_2 (Y_{\mathbf{120}})_{\alpha \beta} (Y_{\mathbf{120}})_{\gamma \delta}+x_3 (Y_{\mathbf{10}})_{\alpha \beta} (Y_{\mathbf{\overline{126}}})_{\gamma \delta} \\ & 
+x_4 (Y_{\mathbf{\overline{126}}})_{\alpha \beta} (Y_{\mathbf{10}})_{\gamma \delta}+x_5 (Y_{\mathbf{\overline{126}}})_{\alpha \beta} (Y_{\mathbf{120}})_{\gamma \delta} \\ &
+x_6 (Y_{\mathbf{120}})_{\alpha \beta} (Y_{\mathbf{\overline{126}}})_{\gamma \delta}+x_7 (Y_{\mathbf{10}})_{\alpha \beta} (Y_{\mathbf{120}})_{\gamma \delta} \\ & 
+x_8 (Y_{\mathbf{120}})_{\alpha \beta} (Y_{\mathbf{10}})_{\gamma \delta}+x_9 (Y_{\mathbf{\overline{126}}})_{\alpha \delta} (Y_{\mathbf{120}})_{\beta \gamma}\\ &
+x_{10} (Y_{\mathbf{120}})_{\alpha \delta} (Y_{\mathbf{120}})_{\beta \gamma} \\ 
C_{\alpha \beta \gamma \delta}^L & =(Y_{\mathbf{10}})_{\alpha \beta} (Y_{\mathbf{10}})_{\gamma \delta}+x_1 (Y_{\mathbf{\overline{126}}})_{\alpha \beta} (Y_{\mathbf{\overline{126}}})_{\gamma \delta}\\ &
-x_3 (Y_{\mathbf{10}})_{\alpha \beta} (Y_{\mathbf{\overline{126}}})_{\gamma \delta} -x_4 (Y_{\mathbf{\overline{126}}})_{\alpha \beta} (Y_{\mathbf{10}})_{\gamma \delta} \\ & 
+y_5 (Y_{\mathbf{\overline{126}}})_{\alpha \beta} (Y_{\mathbf{120}})_{\gamma \delta} +y_7 (Y_{\mathbf{10}})_{\alpha \beta} (Y_{\mathbf{120}})_{\gamma \delta} \\ &
+y_9 (Y_{\mathbf{120}})_{\alpha \gamma} (Y_{\mathbf{\overline{126}}})_{\beta \delta}+y_{10} (Y_{\mathbf{120}})_{\alpha \gamma} (Y_{\mathbf{120}})_{\beta \delta},
\end{aligned}
\ee
where $x_i$ and $y_i$ are the matrices' elements that diagonalise the colour triplet mass matrix and can be treated as free parameters that we vary randomly in the interval $[0,1]$.
In \equaref{eq:effective_superp3} all the Yukawa couplings are rotated to be in the mass basis. 
The coefficients $C^L_{\alpha \beta \gamma \delta}$ and $C^R_{\alpha \beta \gamma \delta}$ induces uncertainties for the partial lifetime. Fitting of fermion masses and mixing helps to obtain $h$, $g$ and $h'$, overall factors between them and $Y_{\bf 10}$, $Y_{\bf 126}$ and $Y_{\bf 120}$, in particular, $V_{11}$ between $h$ and $Y_{\bf 10}$, are undermined. By requiring the perturbativity of the theory, we scan and obtain the maximal and minimal contributions of these coefficients to the proton decay. In Fig.~(4), the exclusion region for mass scales set by Super-K and the future sensitivity of JUNO is obtained by considering the maximal and minimal contribution of coefficients, respectively.

In order to be able to predict proton decay, the squarks or the sleptons in the 5-dimensional operators of \equaref{eq:effective_superp3} need to be dressed with gaugino or higgsino vertices. We note that from the model's symmetry, namely that a bino or a gluino dressing gives zero contribution due to the Fierz identity, only the contributions from the wino and the higgsino dressing significantly promote decay in this channel. In contrast, the dressing from gluino and binos is negligible.
The sub-operators we consider are \cite{Goh:2003nv,Severson:2015dta}
\be
 \begin{aligned}
  C_{\widetilde{W}}^I &= \frac{1}{2}\left(\overline{u^c_L} {d_L}_\beta\right) C_{[\alpha \beta 1] \delta}^L U_{\alpha \alpha^{\prime}}^d U_{\delta \delta^{\prime}}^\nu\left({\overline{d^c_L}}_{\alpha^{\prime}} {\nu_L}_{\delta^{\prime}}\right) \\
  C_{\widetilde{W}}^{IV} &=-\frac{1}{2}\left(\overline{{d^c_L}}_{\beta} {\nu_L}_\delta\right) C_{\alpha[\beta \gamma] \delta}^L U_{1 \alpha^{\prime}}^d U_{\gamma 1}^u\left(\overline{{d^c_L}}_{\alpha^{\prime}} u_L \right) \\
   C_{\bar{h}^{\pm}}^{III} & =-\left(\overline{{d^c_L}}_\beta {\nu_L}_\delta\right) \widehat{C}_{\alpha[\beta \gamma] \delta}^L y_{\alpha \alpha^{\prime}}^{d \dagger} y_{\gamma 1}^{u \dagger}\left(\overline{{d^c_R}}_{\alpha^{\prime}} u_R \right) \\ C_{\bar{h}^{\pm}}^{IV} & =\left(\overline{u^c_R} {{d_R}_\beta}\right) \widehat{C}_{[\alpha \beta 1] \delta}^R y_{\alpha \alpha^{\prime}}^u y_{\delta^{\prime}}^e\left(\overline{{d^c_L}}_{\alpha^{\prime}} {\nu_L}_{\delta^{\prime}}\right) \\
   C_{\bar{h}^{0}}^{I I I} & =-\left(\overline{{d^c_L}}_\beta {\nu_L}_\delta\right) \widehat{C}_{\alpha[\beta \gamma] \delta}^L y_{\alpha \alpha^{\prime}}^{d \dagger} y_{\gamma 1}^{u \dagger}\left(\overline{{d^c_R}}_{\alpha^{\prime}} u_R\right) 
 \end{aligned}
 \label{eq:suboperators}
\ee
where we followed the notation of \cite{Severson:2015dta} and the superscripts label the Feynman diagram, which contributes to the kaon channel, and the subscript label the gaugino, which dresses the Feynman diagram.
Considering the above sub-operators, we note that they are proportional to the mixing matrices and the Yukawa couplings at low energies. Therefore the proton lifetime is tightly linked to the predictions of the scan, and therefore is possible to test the parameter space of the Yukawa sector of the theory with proton decay experiments.
In particular, the free parameters on which the overall lifetime depends are $r_1, r_2,a_1, a_2, c_e, c_\nu$ which are fixed from the scan, $x_i, y_i$, and the Higgs mixing parameters $U_{12}$, $V_{11}$ and $U_{13}$.
We found that by randomly varying $x_i$ and $y_i$, we only achieve a few percent difference in the lifetime while the influence of the Higgs mixing parameters is much more important. The parameter $U_{12}$ can be fixed since it depends on the known parameters $m_0, r_1$, and $M_{B-L}$. The other two parameters are constrained by relations of \equaref{eq:SO10_Yukawas}, and we are considering the highest allowed value. 

We can compute the proton decay width considering two different operators proportional to the sum of the coefficients $\mathscr{C}_{\widetilde{W}}$ and $\mathscr{C}_{h}$, which are called respectively $\mathcal{O}_{\widetilde{W}}$ and $\mathcal{O}_{\bar{h}}$.
These two operators are expressed as:
\bea
\mathcal{O}_{\widetilde{W}}=\left(\frac{i \alpha_2}{4 \pi}\right)\left(\frac{1}{M_T}\right) I\left(M_{\widetilde{W}}, m_{\bar{q}}\right) \mathscr{C}_{\bar{W}}^{\mathcal{A}}\,,\nonumber\\ 
\mathcal{O}_{\bar{h}}=\left(\frac{i}{16 \pi^2}\right)\left(\frac{1}{M_T}\right) I\left(M_{\bar{h}}, m_{\bar{q}}\right) \mathscr{C}_{\bar{h}}^{\mathcal{A}}\,, \label{eq:protondecaywidth1}
\eea 
where $M_T \sim M_{\rm GUT}$ is the mass of the Higgs triplet and $I(a,b)$ is the loop contribute; we have $I(a,b) \simeq a/b^2$ when $a \ll b$.
Finally, the proton decay width can be expressed in terms of these two operators as
\bea
\Gamma(p\rightarrow K^+ \nu ) &=&\frac{M_p}{8\pi}\left( 1- \frac{m_{K^+}^2}{M_p^2}\right) \langle K^+|(u s)_L u_L| p\rangle^2 \nonumber\\
&&\times A_L^2 A_S^2 \left( |\mathcal{O}_{\widetilde{W}}|^2 + |\mathcal{O}_{\bar{h}}|^2 \right)\,,
\label{eq:protondecaywidth2}
\eea
From Eq. \ref{eq:protondecaywidth1} and Eq. \ref{eq:protondecaywidth2}, we can also see that the lifetime is highly dependent on the value of the ratio between the gaugino mass and the SUSY breaking scale and from the SUSY breaking scale itself.

\bibliographystyle{apsrev4-2}
\bibliography{SUSYGUT}

\end{document}